\documentclass{jfm}

\usepackage{custom}
\usepackage[normalem]{ulem}
\usepackage[usenames, dvipsnames]{color}

\graphicspath{{./graphics/}}

\setkeys{Gin}{width=.49\textwidth}

\newcommand{\textcite}[1]{\citet{#1}}
\newcommand{\parencite}[1]{\citep{#1}}

\newcommand{\major}[1]{\textcolor{black}{#1}} 
\newcommand{\minor}[1]{\textcolor{black}{#1}}

\shorttitle{The effective diffusivity of ordered and freely evolving bubbly suspensions}
\shortauthor{A. Loisy, A. Naso, P. D. M. Spelt}

\title{The effective diffusivity of ordered and freely evolving bubbly suspensions}

\author{Aurore Loisy\footnote{Present address: School of Mathematics, University of Bristol, University Walk, Bristol BS8 1TW, United Kingdom.}\corresp{\email{aurore.loisy@bristol.ac.uk}},
  Aurore Naso
 \and Peter D. M. Spelt
}

\affiliation{
Laboratoire de M\'ecanique des Fluides et d'Acoustique,\\
CNRS, Universit\'e Claude Bernard Lyon 1, \'Ecole Centrale de Lyon, INSA de Lyon \\
36 avenue Guy de Collongue, 69134 \'Ecully cedex, France
}

\begin{document}

\maketitle

\begin{abstract}
We investigate the dispersion of a passive scalar such as the concentration of a chemical species, or temperature, in homogeneous bubbly suspensions, by determining an effective diffusivity tensor. Defining the longitudinal and transverse components of this tensor with respect to the direction of averaged bubble rise velocity in a zero mixture velocity frame of reference, we focus on the convective contribution thereof, this being expected to be dominant in commonly encountered bubbly flows. We first extend the theory of \textcite{Koch1989} (which is for dispersion in fixed beds of solid particles under Stokes flow) to account for weak inertial effects in the case of ordered suspensions. In the limits of low and of high P\'eclet number, including inertial effect of the flow does not affect the scaling of the effective diffusivity with respect to the P\'eclet number. These results are confirmed by direct numerical simulations performed in different flow regimes, for spherical or very deformed bubbles and from vanishingly small to moderate values of the Reynolds number. Scalar transport in arrays of freely rising bubbles is considered by us subsequently, using numerical simulations. In this case, the dispersion is found to be convectively enhanced at low P\'eclet number, like in ordered arrays. At high P\'eclet number, the Taylor dispersion \minor{scaling} obtained for ordered configurations is replaced by \minor{the one characterizing} a purely mechanical dispersion, like in random media, even if the level of disorder is very low.
\end{abstract}

\section{Introduction}

Bubble columns are commonly used in a broad range of technologies, notably in the chemical and biochemical industry. Simple bubble columns do not require active stirring and can therefore operate without interior moving parts.
The large surface area between gas and liquid is useful for mass and species transfer, possibly involving chemical reactions \citep[e.g.,][]{Deckwer} such as in air-lift bioreactors, an example of which is in the treatment of wastewater. Bubble columns are also used for this reason in direct contact heat transfer \citep[e.g.,][]{Hewittbook}.
Besides offering a large surface area, rising bubbles agitate the liquid flow, which results in enhanced mixing that usually is desired, but this also poses a modelling difficulty. 
Similar mixing arises also in the diffusion through porous media in the presence of flow, which has been well studied previously, but mostly for fixed beds of particulates, often under creeping flow \citep[e.g.,][]{Batchelor1974,Koch1985}. Mixing in bubble columns is complicated further by the fact that liquid velocity fluctuations are coupled with the dynamics of (deformable) bubbles, usually beyond creeping flow \citep[e.g.,][]{Almeras2015}.

In the present study, we consider transport of a scalar (such as the concentration of a chemical species, or the temperature) through
incompressible bubbly flows.  
Gradients of temperature and concentration may, in general, induce fluid motion and influence the velocity field through changes in density and viscosity, or through the  interface rheology.
If these effects are small, as assumed herein, temperature and solute concentration can be considered as passive scalars. 
Although the arbitrary choice was made in this study to use the terminology of the mass transfer problem, the results carry over to thermal applications  (upon assuming that effects of viscous heating can be ignored).

Our present main interest is the formulation and closure of conservation equations and constitutive relations governing the dispersion of such a scalar in a bubbly suspension over scales (termed hereinafter the ``macroscale'') that are much larger than the bubble size (termed hereinafter the ``microscale'').
Under the assumption of macroscale homogeneity and stationarity, scalar dispersion in multiphase systems can be described by a macroscale version of Fick's (or Fourier's) law which relates the macroscale scalar flux to the macroscale scalar gradient through an effective diffusivity tensor (or effective conductivity tensor in thermal applications) \parencite{Batchelor1974,Koch1985,Koch1987a}. 
This effective diffusivity is defined from an Eulerian perspective.
Experimentally, scalar dispersion is usually investigated from a Lagrangian point of view.
In the Lagrangian framework, the effective diffusivity is defined as the long-time limit of the time rate of change of a fluid tracer's mean-square displacement, that is, as a measure of spread about the mean position.
\textcite{Koch1987a} demonstrated that the Lagrangian effective diffusivity is equivalent to the symmetric part of the Eulerian effective diffusivity, and that the antisymmetric part of the Eulerian effective diffusivity is associated with anisotropic microstructures.

Scalar dispersion in a suspension of particulates (bubbles, drops, or rigid particles) results from two processes of very different nature: the diffusion by Brownian motion of the molecules, and the convection by the fluid velocity disturbances induced by the particulate motion.
The relative importance of these two processes is measured by the P\'eclet number $\Pe = U d_b / D$, where $U$ is the characteristic velocity of the particulates relative to that of the system (defined in \cref{sec:problem_formulation}), $d_b$ is the characteristic size of the particulates, and $D$ is the diffusivity of the bulk.
In the limit $\Pe = 0$, the effective diffusivity is purely diffusive and depends only on the particulate-to-bulk diffusivity ratio, possible discontinuity of the scalar at the interface, particulate volume fraction, and suspension microstructure (i.e., the positions, shapes, and orientations of the inclusions).
This particular situation is essentially relevant to heat and electricity conduction in composite materials.
When $\Pe \gg 1$, the dominant contribution to the effective diffusivity is due to convective mixing.
This last regime is that generally encountered in bubbly flows.

Recently \textcite{Almeras2015} investigated experimentally the dispersion of a low-diffusive dye within a homogeneous swarm of high-Reynolds-number rising bubbles at $\Pe = \O(10^6)$; herein we define the Reynolds number as $\Re = U d_b/\nu_c$ where $\nu_c$ is the kinematic viscosity of the liquid.
They showed that scalar mixing primarily results from pseudo-turbulence, i.e., from the liquid agitation produced by bubble wake interactions, and can be modeled in a manner analogous to dispersion in shear-induced turbulence \parencite{Taylor1922}.
Apart from the work of \textcite{Almeras2015}, the only other experimental investigation of mixing in homogeneous bubbly flows reported in the literature is the preliminary study of \textcite{Mareuge1995} which consists in a single data point.
To the best of our knowledge, neither theoretical nor numerical investigations of scalar mixing in homogeneous bubbly flows have been reported thus far.
Theoretical work is, however, available for other types of multiphase systems, and we shall review these now.

The determination of such an effective diffusivity, at the macroscale, necessitates consideration of the conditions at the microscale. One class of analytical work is devoted to the study of dilute systems with fixed random microstructure, for instance, as a model of a porous medium.
In the absence of convection ($\Pe = 0$), the analytical expression of the effective diffusivity is available in the dilute limit from analysis of the corresponding problem in conduction of heat or electricity through a dispersed medium (e.g., \textcite{Maxwell1873book}, \textcite{Jeffrey1973}). 
The problem of scalar dispersion in the presence of a bulk convective motion ($\Pe > 0$) has been analyzed by \textcite{Koch1985} for Stokes flow through a random bed of fixed solid spheres.
Using the method of conditional averaging pursued earlier by \textcite{Hinch1977}, they carried out an asymptotic analysis in low volume fraction of the effective diffusivity for all values of the P\'eclet number.
Three mechanisms causing dispersion at high P\'eclet number were identified: mechanical dispersion resulting from the stochastic velocity field in the bulk, which is independent of Brownian diffusion and grows as $U d_b$, holdup dispersion in stagnant and recirculating regions which is proportional to $U^2 d_b^2 / D$, and boundary-layer dispersion which grows as $U d_b \ln (U d_b / D)$ near the solid particle surfaces.

Another class of analytical studies assumes a periodic microstructure.
For the pure diffusion problem ($\Pe = 0$), analytical solutions have been derived for a composite material consisting of regularly arranged spheres embedded in a homogeneous matrix \parencite{Rayleigh1892,Sangani1983}, and the effect of anisotropy has been investigated by considering periodic arrangements of spheroidal inclusions \parencite{Kushch1997,Harfield1999}.
In the presence of convection ($\Pe > 0$), the general theory of dispersion developed by \textcite{Brenner1980} and \textcite{Brenner1982} provides a consistent framework for determining the effective diffusivity in spatially periodic media.
\textcite{Koch1989} carried out explicit calculations for a periodic porous medium consisting of fixed solid particles arranged in a cubic lattice and embedded in a continuous phase under Stokes flow conditions.
They showed that in ordered systems, the mechanical dispersion encountered in random media is absent, and that at high P\'eclet number, either Taylor dispersion, growing as $U^2 d_b^2 / D$, or enhanced diffusion, which is proportional to $D$, is obtained depending on the direction of the mean flow relative to the lattice structure.

In bubbly flows, the spatial arrangement of the inclusions evolves in time, the microstructure of the suspension is unknown a priori, and Stokes flow is usually not applicable.
For these reasons, prior analyses are, a priori, not applicable to bubbly suspensions.
Nevertheless, we showed in prior work \citep{Loisy2017b} that the dynamics of freely evolving bubbly suspensions at moderate Reynolds number shares some common features with that of ordered arrays of bubbles.
It is therefore of fundamental interest to investigate, contrast and compare the mixing properties of ordered and freely evolving bubbly suspensions in light of prior asymptotic analyses for ordered and random arrangements of rigid particles.

In this paper we investigate scalar dispersion, by determining the effective diffusivity in ordered and freely evolving bubbly suspensions, specifically, the contribution of bubble-induced velocity disturbances thereof. The prior work outlined above has established that in the systems studied therein, the effective diffusivity can be much larger than that in each of the fluids involved, even if the diffusivity in the two media is the same and the scalar is continuous at the surface of particulates. In view of the already significant number of parameters involved, we shall therefore adopt this restriction here. \minor{Such a simplified approach will not provide an accurate description of real bubbly flows, but \minor{should shed} some light on the fundamental mechanisms of mixing in these systems.}

The paper is organised as follows. The theoretical framework and problem statement are provided in \cref{sec:problem_formulation}. Our numerical approach to compute the effective diffusivity is presented, and followed by a  description of the regimes and the range of parameter values that are investigated herein, in \cref{sec:numerical_methodology}.
The first objective (in \cref{sec:ordered_arrays}) is to elucidate the role played by liquid inertia in ordered suspensions, using direct numerical simulation and analysis.
The second objective (in \cref{sec:free_arrays}) is to investigate the effective diffusivity of freely evolving suspensions for a wide range of P\'eclet numbers, to compare it with that obtained for ordered suspensions, and to evaluate the effect of introducing additional degrees of freedom in the system.
Finally, the main results and perspectives of this work are provided in \cref{sec:conclusion}.

\section{Problem statement}
\label{sec:problem_formulation}


\noindent
The local evolution of the passive scalar $c$ in each fluid is governed by
\begin{subequations}
\label{eq:local_scalar_eq}
\begin{align}
\pd{c}{t} + \div \v{q} & = 0 \label{eq:local_scalar_conservation} \\
\intertext{where $\v{q}$ is the flux of scalar given by} 
\quad \v{q} & = \v{u} c - D \grad c \label{eq:local_scalar_flux_def}
\end{align}
\end{subequations}
with $\v{u}$ the fluid velocity and $D$ the constant scalar diffusivity.
We assume that the scalar and its gradient are continuous across the interface, and phase change is not considered in this study. Under these assumptions, no distinction between the phases is needed for the scalar transport, which is described by \cref{eq:local_scalar_eq} in the entire system.
\major{We return to these restrictions in \cref{sec:macroscale_description} and in the Conclusions section; the objective here is to study this key basic reference problem.}

In the context of heat transfer, \cref{eq:local_scalar_eq} derives from the energy balance upon neglecting viscous heating, in this case $c$ would represent the temperature\major{, continuous at the interface,} and $D$ the thermal diffusivity as defined by Fourier's law, assumed to be equal in both gas and liquid.
In the context of mass transfer, \cref{eq:local_scalar_eq} describes the transport of a chemical species present at very low concentration $c$ so that Fick's law describes the conservation of mass, neglecting any difference in molecular diffusivity $D$ and solubility of the species in the two phases. \major{While the assumption of equal molecular diffusivities is never satisfied in real systems, the assumption of a unit dimensionless Henry's constant is reasonably applicable to, e.g., carbon dioxide dispersion in the air-water system.}

The fluid motion in the gas and liquid is governed by the incompressible Navier-Stokes equations, which are coupled at the interface by the appropriate jump conditions, namely the continuity of velocity and of tangential traction across the interface, and a jump in normal traction due to surface tension.



\subsection{\minor{Macroscale description} \label{sec:macroscale_description}}

The problem we are concerned with here is the modeling of scalar transport at a macroscale, that is, at the scale whereat the suspension may be seen as a homogeneous continuum, without distinction between the two phases.
In order to obtain such a macroscopic description, we consider an ensemble of realizations of the suspension, these realizations having the same macroscopic conditions (e.g., fluid properties, gas volume fraction) but different microscopic configurations (e.g., bubble individual positions, shapes and velocities), and average over those realizations.
In concrete terms, ensemble averaging would be realized by averaging over a large number of experiments run under identical macroscopic conditions.
The ensemble-averaged transport equation is obtained from ensemble averaging the local transport equation \cref{eq:local_scalar_eq}.
It reads
\begin{equation}
	\pd{\avg{c}}{t} + \div \avg{\v{q}} = 0,
	\label{eq:conservation_avg}
\end{equation}
where $\avg{\,}$ denotes the ensemble average operator, and where the ensemble-averaged flux is given by
\begin{equation}
	\avg{\v{q}} = \avg{\v{u}} \avg{c} - D \grad \avg{c} + \avg{ \v{u}' c' }
		\label{eq:def_average_flux}
\end{equation}
where the velocity fluctuations are defined by $\v{u}' = \v{u} - \avg{\v{u}}$ and the scalar fluctuations by $c' = c - \avg{c}$.
Under the restrictions set out above, the average flux consists of three contributions: (i) $\avg{\v{u}} \avg{c}$ is the advection of the average scalar field at the average system velocity; (ii) $- D \grad \avg{c}$ is the diffusion of the average scalar field directly by the average scalar gradient; (iii) $\avg{ \v{u}' c' }$ corresponds to the advection of the scalar fluctuations by the velocity fluctuations induced by bubble motion. 

When the suspension is statistically homogeneous and in a statistically stationary state, the linearity in $c$ of the local flux \cref{eq:local_scalar_flux_def} results, in the presence of an imposed constant average scalar gradient, in a macroscale constitutive relation of the form \parencite{Koch1985,Koch1987a}:
\begin{equation}
	\avg{\v{q}} = \avg{\v{u}} \avg{c} - \m{D}^{\mathsf{eff}} \bcdot \grad \avg{c}
	\label{eq:def_effdiff}
\end{equation}
where $\m{D}^{\mathsf{eff}}$ is a constant effective diffusivity tensor.
Comparison of the effective diffusivity definition \cref{eq:def_effdiff} with the average flux expression \cref{eq:def_average_flux} yields the expression of the effective diffusivity.
In order to reflect the contributions to the scalar flux identified above, it is customary to write the effective diffusivity as
\begin{equation}
	\m{D}^{\mathsf{eff}} = D \m{I} + \m{D}^{\mathsf{conv}}
	\label{eq:Deff_expr}
\end{equation}
where
\begin{equation}
	\m{D}^{\mathsf{conv}} \bcdot \grad \avg{c} = - \avg{ \v{u}' c' } 
	\label{eq:Dconv_expr}
\end{equation}
is the convective contribution arising from bubble-induced velocity fluctuations. 
For this model to be complete, one must find a closure relation for $\m{D}^{\mathsf{conv}}$ only in terms of macroscopic quantities appearing in the problem statement. 
We recall here that further contributions to the average flux \cref{eq:def_average_flux} and hence to the effective diffusivity \cref{eq:Deff_expr} arise if the diffusivity in the fluids are not the same, or if the concentration is discontinuous at fluid/fluid interfaces (e.g., \textcite{Batchelor1977,Koch1985}). 
We return to the significance of this in \cref{sec:conclusion} below.

\subsection{\minor{Effective transport properties}}

To determine the effective diffusivity for (unbounded) homogeneous bubbly suspensions, we represent such flows by the periodic repetition of a cubic unit cell containing a finite number $N_b$ of freely moving bubbles of equal volume,
building on our prior work on the dynamics of bubbles for this model system \citep{Loisy2017b}.
In the limit $N_b = 1$, one obtains a simple cubic array of bubbles, which is of interest as a model of perfectly ordered suspensions.
The opposite limit of large $N_b$ is of interest as a model of real suspensions, although convergence with the number of bubbles would have to be verified.
We shall refer hereinafter to this setup with one bubble in the cell as an {\it ordered} array, and to that with more than one bubble in the unit cell as a {\it free} array.

The bubbles rise under the sole effect of buoyancy. Herein, 
\minor{an upward-pointing primary axis $\v{e}_3$ of the periodic arrangement is taken to be aligned with gravity  (with the exception of the more general analysis presented in \cref{subsec:asymptotic_analysis}).}
From symmetry arguments, and adopting a Cartesian coordinate system,
\begin{equation}
	\m{D}^{\mathsf{conv}} = \left[ \begin{matrix}
							D^{\mathsf{conv}}_{\perp} 	& D^{\mathsf{conv}}_{12} 	& D^{\mathsf{conv}}_{13} \\
							D^{\mathsf{conv}}_{12} 		& D^{\mathsf{conv}}_{\perp} & D^{\mathsf{conv}}_{13} \\
							D^{\mathsf{conv}}_{31} 		& D^{\mathsf{conv}}_{31} 	& D^{\mathsf{conv}}_{\parallel} \\
							\end{matrix} \right]
\end{equation}
where we have introduced the longitudinal and transverse components of the convective contribution to the effective diffusivity, denoted $D^{\mathsf{conv}}_\parallel$ and $D^{\mathsf{conv}}_\perp$, respectively, and defined by
\begin{equation}
	D^{\mathsf{conv}}_\parallel = D^{\mathsf{conv}}_{33} \qquad \mbox{and} \qquad D^{\mathsf{conv}}_\perp = D^{\mathsf{conv}}_{11} = D^{\mathsf{conv}}_{22}.
\end{equation}

Our first goal is to characterize the effects of liquid inertia (through $\Re$) on the dependence of $\m{D}^{\mathsf{conv}}$ on $\Pe$ for ordered suspensions ($N_b = 1$), thereby extending prior work on dilute ordered arrays of rigid spheres in Stokes flow conditions \parencite{Koch1989}.
Our second goal is to evaluate the effect of introducing additional degrees of freedom in the system (through increasing $N_b$), and to investigate the dependence of $\m{D}^{\mathsf{conv}}$ on $\Pe$ in freely evolving suspensions (sufficiently large $N_b$).
As we found the off-diagonal components to be zero in all configurations that we investigated, only results for the longitudinal and the transverse components of $\m{D}^{\mathsf{conv}}$ will be presented.

In dimensionless groups, we shall use as characteristic length scale the bubble size $d_b$, which is defined, since bubbles are deformable, as the (equivalent) diameter of a sphere of the same volume.
The characteristic velocity $U$ is taken here as the bubble rise velocity in the frame of the suspension (the so-called drift velocity $\avg{\v{U}}=\avg{\v{u}}_d - \avg{\v{u}}$, where the first term is the volume average of velocity on the disperse phase only and the second one is the same average in the entire system).
As already mentioned, a key dimensionless group appearing in the scalar transport problem is the P\'eclet number $\Pe = U d_b/D$ which compares advective and diffusive transport. 
Our main objective is to elucidate the effect of the value of $\Pe$ on the effective diffusivity using analytical and numerical methods. 

The effective diffusivity necessarily also depends on the gas volume fraction $\phi= (N_b \upi d_b^3) / ( 6 h^3)$ ($h$ is the linear size of the unit cell); the analytical and computational methods used here pose some restrictions on the range of $\phi$ values that can be studied herein, we postpone discussion of that to the pertinent sections below.
We also consider the effects of the number of bubbles in the periodic cell, $N_b$, which affects the order in the suspension: $N_b=1$ corresponds to a cubic array, whereas more bubbles results in a different microstructure (the latter term encompasses all the information about the statistical distribution of the bubble positions, shapes, orientations, etc.).
Since scalar transport is coupled to momentum transport, the bubble Reynolds number $\Re = U d_b/\nu_c$ may also play a significant role that will be investigated here as well. 
The ranges of $\phi$, $N_b$ and $\Re$ studied here are summarized in \cref{tab:cases_parameters_sublist}.

As the bubbly flows we consider are buoyancy-driven, a difficulty arises from the fact that $U$ is a priori unknown, and depends in a complex manner on  $N_b$, $\phi$, the density and viscosity ratios between both phases, the  Archimedes (or Galileo) number $\Ar = \sqrt{\rho_c \abs{\rho_d-\rho_c} g d_b^3} / \mu_c$, and the Bond (or E\"otv\"os) number $\Bo = \abs{\rho_d-\rho_c} g d_b^2 / \gamma$, where the subscripts $d$ and $c$ refer to the disperse (gas) and continuous (liquid) phases, respectively,  $g$ is the magnitude of the gravitational acceleration, $\rho$ denotes density, $\mu$ is the dynamic viscosity, and $\gamma$ is the surface tension.
In most bubbly flows of practical relevance, the gas-to-liquid density and viscosity ratios are vanishingly small.
Their precise values are not important from a physical point of view as long as they are small enough; in the simulations, the gas-to-liquid density and viscosity ratios were set to $\rho_d/\rho_c = 10^{-3}$ and $\mu_d/\mu_c = 10^{-2}$, respectively.
The dependence of $U$ on ($\Ar$, $\Bo$, $\phi$, $N_b$) has been addressed in \cite{Loisy2017b} and is not further discussed here.
In the present study, we shall therefore assume that $U$ is known.

\section{Methodology}
\label{sec:numerical_methodology}

\noindent
For convenience of numerical implementation, we reorganise the problem formulation by introducing the decomposition
\begin{equation}
	c = \bar{c} + \tilde{c}
	\label{eq:scalar_alternative_decomposition}
\end{equation}
where $\bar{c}$ is the imposed constant linear scalar field
\begin{equation}
	\bar{c} = \grad \avg{c}\cdot \v{x}. 
	\label{eq:def_cbar}
\end{equation}
The advantage of this decomposition is that the disturbance field $\tilde{c}$ is then spatially periodic.
The governing equation for this disturbance field is
\begin{equation}
	\pd{\tilde{c}}{t} + \div (\v{u} \tilde{c}) - \div ( D \grad \tilde{c} ) = - \v{u} \bcdot \grad \avg{c} 
	\label{eq:scalar_eq_numerical}
\end{equation}
which is the equation we integrate numerically.
The convective contribution to the effective diffusivity is then calculated from
\begin{equation}
	\m{D}^{\mathsf{conv}} \bcdot \grad \avg{c} = - \avg{\v{u}' \tilde{c}} \label{eq:Dconv_practical}
\end{equation}
which can be shown to be equivalent to \cref{eq:Dconv_expr}.
In this expression, $\avg{\;}$ is defined as an ensemble average operator, as above.
For statistically homogeneous and stationary systems, as considered here, it is inferred from ergodicity that ensemble averaging is identical to volume and time averaging.
As a consequence, $\m{D}^{\mathsf{conv}}$ is computed from \cref{eq:Dconv_practical} with the ensemble average being replaced in practice by a volume average combined with a time average over an appropriate time period.

\subsection{Numerical method}
\label{sec:numerical_method}

Thus, the components of $\m{D}^{\mathsf{conv}}$ are obtained from direct numerical simulations (DNS) by imposing a constant linear scalar field $\bar{c}$ and determining the resulting periodic disturbance scalar field.
Two distinct simulations are required to fully determine the five independent components of $\m{D}^{\mathsf{conv}}$: in one simulation, $\grad \bar{c} = \v{e}_3$, which yields $D^{\mathsf{conv}}_{13}$ and $D^{\mathsf{conv}}_{\parallel}$, in the other simulation, $\grad \bar{c} = \v{e}_1$, which yields $D^{\mathsf{conv}}_{\perp}$, $D^{\mathsf{conv}}_{12}$, and $D^{\mathsf{conv}}_{31}$.
The off-diagonal components of $\m{D}^{\mathsf{conv}}$ were found to be zero (up to computer accuracy for ordered arrays, and statistical uncertainty for free arrays) for all the sets of parameters we considered, and therefore will not be shown.

The numerical methods employed to solve the two-phase flow have been described in detail in \textcite{Loisy2017b}.
In short, we employ a standard projection method \parencite{Chorin1968} to integrate the incompressible Navier-Stokes equations, a level-set method (e.g., \parencite{Sussman1994}) to capture the moving gas-liquid interface, and surface tension is accounted for using the continuum surface force model \parencite{Brackbill1992}.

Our algorithm proceeds iteratively through the following steps:
\begin{enumerate}
\item The position of the interface is first advanced in time according to the modified level-set method of \textcite{Sabelnikov2014} using a third-order total-variation-diminishing (TVD) Runge-Kutta scheme. The level-set function is then reinitialized using the procedure of \textcite{Russo2000}, and a correction is finally applied to enforce volume conservation. 
\item The scalar transport equation \cref{eq:scalar_eq_numerical} is advanced by using a mixed Crank-Nicolson/third-order Adams-Bashforth time-stepping scheme.
\item The time integration of the incompressible Navier-Stokes equations is then carried out using a mixed Crank-Nicolson/third-order Adams-Bashforth scheme and consists in the combination of a predictor step, where a temporary velocity field is estimated by ignoring the effect of pressure, and of a corrector step, where the velocity field is corrected by the pressure gradient term computed from the divergence-free condition.
\end{enumerate}

Spatial discretization relies on a mixed finite difference/finite volume approach on a fixed, staggered, Cartesian grid. 
Second-order centered schemes are generally employed, except for advective terms which are discretized using fifth-order weighted-essentially-nonoscillatory (WENO) schemes.

Results of numerical tests are presented in the Appendix.

\subsection{Parametric study}
	
\begin{table}
\centering
\begin{tabular}{lllllll}
	case & $\Bo$ & $\Ar$ & $N_b$    & $\phi$ & $\Re$ & bubble shape 			\\[0.5em]
	S0	& 0.38	& 0.15	& 1        & 0.002 & 0.00164 & spherical 			\\ 
	S1	& 0.38	& 5.03	& 1        & 0.002 & 1.72 	& spherical 			\\ 
	C	& 243	& 15.2	& 1        & 0.002 & 9.44 	& skirted 		\\ 
	E1	& 2.0	& 29.9	& 1 	   & 0.002 & 39.9   & ellipsoidal 			\\
	E1	& 2.0	& 29.9	& $[1,12]$ & 0.024 & $\approx 30$  & ellipsoidal 	 \\ 
\end{tabular}                                                                                                                                                                
\caption{Simulated flow configurations: $\Bo$ and $\Ar$ define the flow regime, $N_b$ is the number of free bubbles in the unit cell, $\phi$ is the gas volume fraction. The resulting bubble Reynolds number ($\Re$) and shape are also provided.
}
\label{tab:cases_parameters_sublist}
\end{table}

Four different flow regimes, as defined by the set ($\Ar$, $\Bo$), are considered here.
These are described in \cref{tab:cases_parameters_sublist}, and have been studied in \cite{Loisy2017b} (the same case code names are used).
In case S0, the bubbles are spherical and the Reynolds number is vanishingly small, which \minor{approaches Stokes flow} conditions. 
In case S1, the bubbles are (nearly) spherical and $\Re\gtrsim 1$.
In case C, the bubbles are skirted, and $\Re \approx 8$.
In case E1, the bubbles are ellipsoidal, and $\Re\approx 30-40$.

Ordered arrays of bubbles in these four flow regimes have been considered for the smallest volume fraction numerically accessible (value provided in \cref{tab:cases_parameters_sublist}).
After a transient regime, all ordered suspensions considered here are in a strictly steady state (for the flow and the scalar) during which the results presented in \cref{sec:ordered_arrays} were obtained.
Simulations of scalar transport in free arrays have been performed for $2 \leqslant N_b \leqslant 12$ in case E1 at $\phi = 2.4$~\%.
In these conditions, coalescence is indeed absent (it does occur at larger $\phi$), whereas simulations at lower $\phi$ for free arrays are excessively expensive for the method and facilities used.
In this regime, the system is in an unsteady but statistically stationary state (for the flow and the scalar), during which the statistics presented in \cref{sec:free_arrays} have been measured.
For each of these configurations ($\Ar$, $\Bo$, $\phi$, $N_b$), the drift velocity (and thereby the Reynolds number) is known from \cite{Loisy2017b}.
This allowed us to impose the P\'eclet number a priori.

The numerical simulation results for ordered arrays are compared with the results of analysis at small (but possibly finite) Reynolds number and small volume fraction.

\section{Ordered suspensions}
\label{sec:ordered_arrays}

\noindent
We examine in this section the dispersion of a passive scalar in ordered suspensions of deformable bubbles. 
Our main objective here is to elucidate the effects of inertia on dispersion, using theoretical analysis and numerical simulation.

\subsection{Asymptotic analysis}
\label{subsec:asymptotic_analysis}

We first determine analytically the convective contribution to the effective diffusivity of ordered suspensions of spherical fluid particulates (bubbles or drops).
The Reynolds number of the particulates is assumed to be small so that the Navier-Stokes equations can be approximated by the Oseen equations.

\subsubsection{General solution}

An ordered array of particulates translating at a drift velocity $\v{U}$ is equivalent to an ordered array of fixed particulates immersed in a viscous fluid moving with an average system velocity $\avg{\v{u}} = -\v{U}$.
The centers of the particulates are located on the nodes of a simple cubic lattice:
\begin{equation}
	\v{r}_{\boldsymbol{n}} = h \, ( n_1 \v{e}_1 + n_2 \v{e}_2 + n_3 \v{e}_3) \qquad n_1,n_2,n_3 = 0, \pm 1, \pm 2, \dots
\end{equation}
where $h$ is the lattice spacing and $\v{e}_i$ are the unit vectors aligned with the primitive axes of the cubic lattice.
In the dilute limit ($d_b / h \ll 1$), the action of these particulates on the fluid can be represented by point forces $-\v{f}$.
The convective contribution to the effective diffusivity arising from the far field has been derived by \textcite{Koch1989} for an ordered array of rigid spheres in the Stokes flow regime.
In what follows we extend their result to 
\minor{the case} of spherical fluid particulates at small but finite $\Re$.

When $\Pe \ll 1$, the convective contribution to the effective diffusivity arising from the far field can be approximated by \parencite{Koch1989}:
\begin{equation}
	\frac{\m{D}^{\mathsf{conv}}}{D} = \sum_{\v{k} \ne \v{0}} \frac
	{k^2 \hat{\v{u}}' (\v{k}) \hat{\v{u}}' (- \v{k})}
	{(2 \upi)^2 k^4 D^2 + ( \v{U} \bcdot \v{k} )^2}, 
	\label{eq:general_expression_Dconv}
\end{equation}
where the summation is over all vectors $\v{k}$ in the reciprocal lattice
\begin{equation}
	\v{k} = \frac{1}{h} \, (n_1 \v{e}_1 + n_2 \v{e}_2 + n_3 \v{e}_3)
\end{equation}
and where $\hat{\v{u}}'$ is the three-dimensional Fourier transform of the velocity disturbance $\v{u}' = \v{u} - \avg{\v{u}}$.
In Oseen flow past an ordered array of point particulates, $\hat{\v{u}}'$ is given by 
\begin{equation}
	\hat{\v{u}}'(\v{k}) = \frac{ \v{f} \bcdot ( \v{k} \v{k}/k^2 - \m{I} ) }{ (2 \upi k)^2 h^3 \mu_c + \i 2 \upi h^3 \rho_c \v{U} \bcdot \v{k} } \qquad \v{k} \ne \v{0},
\end{equation}
where $\v{f}$ is the hydrodynamic force exerted by the ambient fluid on a particulate.
In the dilute limit, $\v{f}$ can be approximated by the Oseen drag exerted on a single spherical \minor{fluid} particulate:
\begin{equation}
	\v{f} = F \v{f}_{0,\mathrm{Stokes}} 
\end{equation}
where $\v{f}_{0,\mathrm{Stokes}}$ is the Stokes drag on 
\minor{that} particulate \parencite{Hadamard1911,Rybczynsky1911}:
\begin{equation}
	\v{f}_{0,\mathrm{Stokes}} = -2 \upi \mu^* \mu_c d_b \v{U}, \qquad \mbox{with\ } \mu^* = \frac{\mu_c + 3 \mu_d/2}{\mu_c + \mu_d},
	\label{eq:drag_Hadamard_Rybczynsky_inDeff}
\end{equation}
and where $F$ accounts for the finite-$\Re$ correction to the Stokes drag \parencite{Brenner1963a}:
\begin{equation}
	F = 1 + \frac{1}{8} \mu^* \Re.
	\label{eq:def_F_Oseen_isolated}
\end{equation}

The convective contribution to the effective diffusivity of a dilute ordered array of fluid particulates in Oseen-flow conditions is therefore:
\begin{subequations}
\label{eq:Deff_analytical}
\begin{equation}
	\frac{\m{D}^{\mathsf{conv}}}{D} = \frac{\mu^{*2}}{(2 \upi)^2} \frac{d_b^2}{h^2} F^2 \m{C}, 
	\label{eq:def_D_tensor}
\end{equation}
where $\m{C}$ is the dimensionless tensor:
\begin{equation}
	\m{C} = \sum_{\v{k}^* \ne \v{0}} \frac{\sq[4]{ \v{U}^* \bcdot \pa[3]{ \dfrac{\v{k}^* \v{k}^*}{k^{*2}} - \m{I} } }^2 }{ k^{*2} \sq[4]{ \dfrac{(2 \upi)^2 k^{*4}}{\Pe_h^2} + ( \v{U}^* \bcdot \v{k}^* )^2} \sq[4]{ 1 + \dfrac{\Re_h^2 ( \v{U}^* \bcdot \v{k}^* )^2}{(2 \upi)^2 k^{*4}} }}
	\label{eq:def_C_tensor}
\end{equation}
\end{subequations}
with $\v{U}^* = \v{U}/U$, $\v{k}^* = \v{k} h$, $\Re_h = \rho_c U h/\mu_c$, and $\Pe_h = U h /D$.
The solution given by \textcite{Koch1989} (equation (4.5) therein) for rigid spheres and Stokes flow is recovered in the limit $\Re \rightarrow 0$ and $\mu_d/\mu_c \rightarrow \infty$.

The tensor $\m{C}$ only depends on $\Pe_h$, $\Re_h$, and on the orientation of $\v{U}$ relative to the reciprocal lattice (which structure is, for cubic arrays, identical to that of the direct lattice).
As highlighted by \textcite{Koch1989}, the asymptotic behavior of $\m{C}$\minor{, and hence of $\m{D}^{\mathsf{conv}}$,} depends on whether there exists any $\v{k}$ such that $\v{U} \bcdot \v{k} = 0$, that is, on whether there exists any separation vector $\v{r}_{\boldsymbol{n}}$ in the real space which is perpendicular to $\v{U}$.
The asymptotic behavior of $\normtwo{\m{D}^{\mathsf{conv}}}$, where $\normtwo{\;}$ denotes the tensorial Frobenius norm, is provided in \cref{tab:asymptotic_Deff_with_inertia}.
The results show that the dependence of $\normtwo{\m{D}^{\mathsf{conv}}}$ on $\Pe$ in the limits \minor{$\Pe_h \ll 1$} and \minor{$\Pe_h \gg 1$} is, qualitatively, not affected by (weak) inertial effects.

\begin{table}
\centering
\begin{tabular}{ccccc}
\multicolumn{2}{c}{regime} &&  \multicolumn{2}{c}{$\normtwo{\m{D}^{\mathsf{conv}}}/(D F^2 d_b^2/h^2)$} \\[0.5em]
$\Pe_h = U h / D$ & $\Re_h = \rho_c U h / \mu_c$ && if $\exists \v{r}_{\boldsymbol{n}} \mid \v{U} \perp \v{r}_{\boldsymbol{n}}$ & if $\nexists \v{r}_{\boldsymbol{n}} \mid \v{U} \perp \v{r}_{\boldsymbol{n}}$ \\[0.5em]
\multirow{2}{*}{$\Pe_h \ll 1$} 	& $\Re_h \ll 1$ && $\Pe_h^2$	& $\Pe_h^2$ \\
								& $\Re_h \gg 1$ && $\Pe_h^2$	& $\Pe_h^2/\Re_h^2$ \\[0.5em]
\multirow{2}{*}{$\Pe_h \gg 1$} 	& $\Re_h \ll 1$ && $\Pe_h^2$	& 1 \\
								& $\Re_h \gg 1$ && $\Pe_h^2$	& $1/\Re_h^2$ \\
\end{tabular}                                                                                                                                                                
\caption{
Asymptotic order of $\normtwo{\m{D}^{\mathsf{conv}}}$ depending on \minor{$\Pe_h$, $\Re_h$}, and on the orientation of the mean flow relative to the real lattice, based on the solution \cref{eq:Deff_analytical}, derived for an ordered array of point particulates in Oseen flow conditions ($F$ is the Oseen drag divided by the Stokes drag).
}
\label{tab:asymptotic_Deff_with_inertia}
\end{table}

\subsubsection{Application to ordered arrays rising vertically}

Let us now come back to our original problem of an ordered array of particulates rising under the effect of buoyancy.
The gravitational acceleration is oriented along a primary axis of the array, $\v{g} = -g \v{e}_3$, and although this is not the only possible solution (see, e.g., \textcite{Loisy2017b}), \minor{we restrict the analysis to} the simplest case of bubbles rising vertically.
In this case the hydrodynamic force exerted by the fluid on a particulate is parallel to the drift velocity, and, since this force balances the buoyancy force at steady state, $F$ is related to $U$ through
\begin{equation}
	F = \frac{U_{0,\mathrm{Stokes}}}{U}
	\label{eq:def_F_with_U}
\end{equation}
where $U_{0,\mathrm{Stokes}}$ is the terminal velocity of an isolated spherical fluid particulate in Stokes flow:
\begin{equation}
	U_{0,\mathrm{Stokes}} = \frac{1}{12} \frac{\abs{\rho_c - \rho_d} g d_b^2}{\mu^* \mu_c}, \qquad \mbox{with\ } \mu^* = \frac{\mu_c + 3 \mu_d/2}{\mu_c + \mu_d}.
	\label{eq:terminal_velocity_Stokes_forDeff}
\end{equation}
Note that $F$ can also be expressed in terms of commonly employed dimensionless groups:
\begin{equation}
	F = \frac{1}{12 \mu^*}\frac{\Ar^2}{\Re}.
	\label{eq:def_F_with_Ar_Re}
\end{equation}
In the ``sedimentation'' problem considered here, $F$ is generally not known (as $U$ is generally not known): it is a non-trivial function of the flow regime and volume fraction which reduces to \cref{eq:def_F_Oseen_isolated} when $\phi \rightarrow 0$ and when Oseen-flow approximation is applicable.

\begin{figure}
\centering
\includegraphics{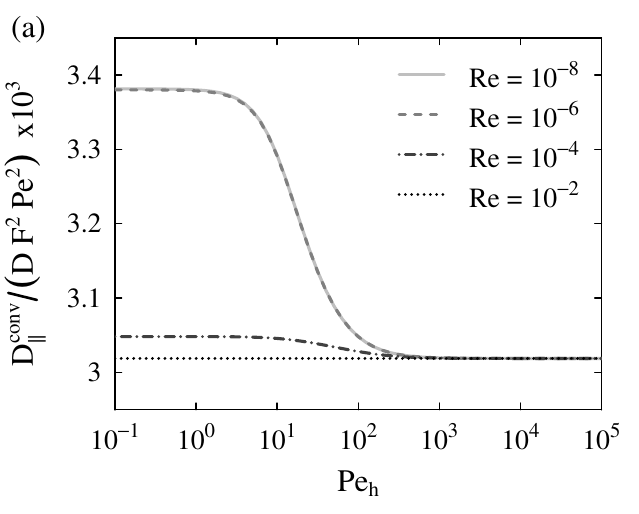}
\includegraphics{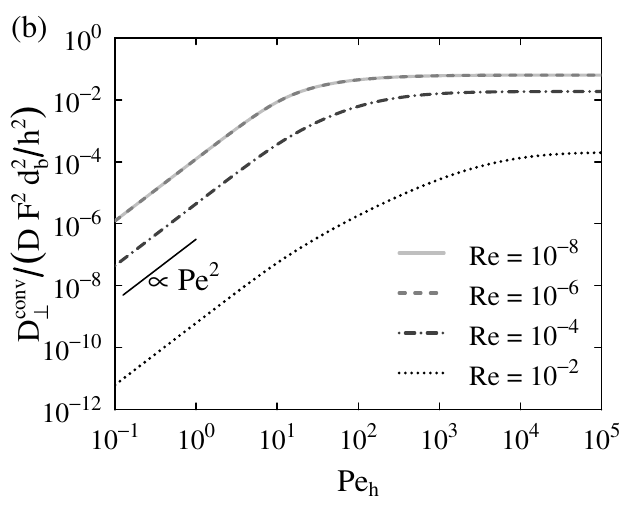}
\caption{Longitudinal (a) and transverse (b) components of $\m{D}^{\mathsf{conv}}$ as a function of the P\'eclet number \minor{based on the lattice spacing ($\Pe_h = U h / D$)} for ordered arrays of point particulates at various small but finite Reynolds numbers ($\v{U} = U \v{e_3}$, $d_b/h = 10^{-6}$, and $F$ is given by \cref{eq:def_F_with_U}).  \minor{Note that in (a), $D^{\mathsf{conv}}_\parallel$ is compensated by $\Pe^2$.}
}
\label{fig:analytical_solution_vs_Pe}
\end{figure}

The longitudinal and transverse components of the convective contribution, $D^{\mathsf{conv}}_\parallel$ and $D^{\mathsf{conv}}_\perp$ respectively, have been calculated from \cref{eq:Deff_analytical} for $d_b/h = 10^{-6}$ as a function of \minor{$\Pe_h$} for various $\Re < 1$.
This very low value of $d_b/h$ is required to allow \minor{$\Pe_h \gg 1$} while satisfying the condition \minor{$\Pe = \Pe_h \, d_b/h \ll 1$} under which the analytical solution has been derived.
The results, shown in \cref{fig:analytical_solution_vs_Pe}, indicate that the asymptotic dependences of $D^{\mathsf{conv}}_{\parallel}$ and $D^{\mathsf{conv}}_{\perp}$ on $\Pe$ are independent of $\Re$.
The sole effect of inertia is to modify the proportionality constants (by a substantial amount for the transverse component though).

In the limit of low \minor{$\Pe_h$ (say, $\Pe_h < 10^{1}$)}, both the transverse and the longitudinal components of $\m{D}^{\mathsf{conv}}$ exhibit a quadratic dependence on the P\'eclet number ($D^{\mathsf{conv}}_{\perp,\parallel} \propto D \Pe^2$). 
In this regime, diffusion is much faster than convection.
As the scalar is advected by velocity disturbances, it rapidly spreads out owing to diffusion, and convective dispersion (measured through $\m{D}^{\mathsf{conv}}$) is influenced by both mechanisms.
This regime corresponds to the ``convectively enhanced dispersion'' regime in \textcite{Koch1989}.

In the limit of high \minor{$\Pe_h$ (say, $\Pe_h > 10^{3}$)}, the transverse component of $\m{D}^{\mathsf{conv}}$ is independent of the P\'eclet number ($D^{\mathsf{conv}}_\perp \propto D$) whereas its longitudinal component grows quadratically with the P\'eclet number ($D^{\mathsf{conv}}_\parallel \propto D \Pe^2$).
In this regime, convection dominates, but owing to the spatial periodicity of the flow, convective dispersion is obtained only if molecular diffusion across streamlines is considered \parencite{Koch1989}.
This regime is termed ``Taylor dispersion'' owing to the formal analogy, pointed out by \textcite{Brenner1980}, with one-dimensional shear-induced Taylor dispersion in a capillary tube.
	
We emphasize that the expression \cref{eq:Deff_analytical} has been derived from the approximation \cref{eq:general_expression_Dconv}, the validity of which is established only for $\Pe \ll 1$ (which is, in practice, of limited use).
Using symmetry arguments, \textcite{Koch1989} (section 4.2 therein) showed that in the limit $\Pe \gg 1$, Taylor dispersion is obtained if the average flow is perpendicular to a set of planes of both translational and reflectional symmetry, such as Stokes flows parallel to the primary axis of an ordered array of spheres.
Taylor dispersion is then easily understood by remarking that, owing to the symmetries of the flow, a fluid tracer particle entering the unit cell at one point, say $\v{x}$, exits the cell at the equivalent point in the next cell, that is, $\v{x} + h \v{e}_3$, so that dispersion can only occur if diffusion across streamlines is present \parencite{Koch1989}.
In the presence of inertial effects, the reflectional symmetry is lost, hence this argument does not hold.
\textcite{Koch1989} also demonstrated that, for Stokes flow, the solution for $\Pe \ll h/d_b$ is identical, at lowest order, to that obtained for $\Pe \ll 1$ (section 4.3 therein, note that their $\Pe$ corresponds to $\Pe_h$ in our notations).
Such a demonstration for Oseen flow will not be attempted here.
Instead, the range $\Pe \geqslant 1$ will be explored using direct numerical simulations.

\subsection{Numerical results}

The above analysis provides explicit expressions of $D^{\mathsf{conv}}_\parallel$ and $D^{\mathsf{conv}}_\perp$.
These are valid for spherical bubbles rising at $\Re < 1$ (strictly speaking, at a Reynolds number sufficiently small to assume Oseen flow, in terms of Archimedes and Bond numbers this regime would be reached for $\Bo < 1$ and $\Ar \lesssim 1$), and in the limits $\phi \rightarrow 0$ and $\Pe \ll 1$.
We shall now determine using numerical simulations whether these restrictions can be relaxed, and if so, to which extent.

\begin{figure}
\centering
\includegraphics{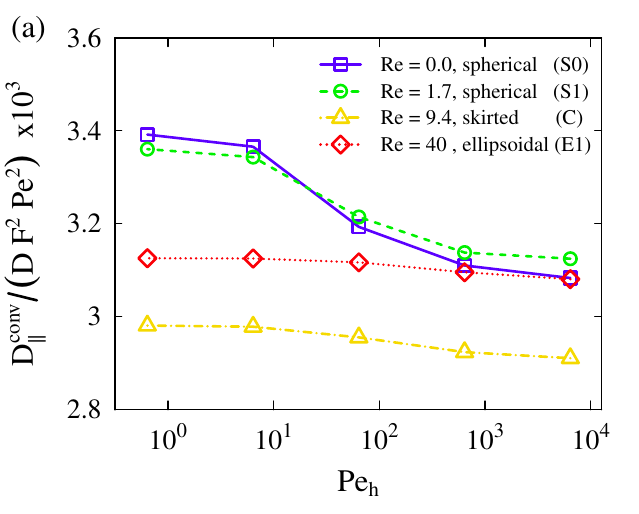}
\includegraphics{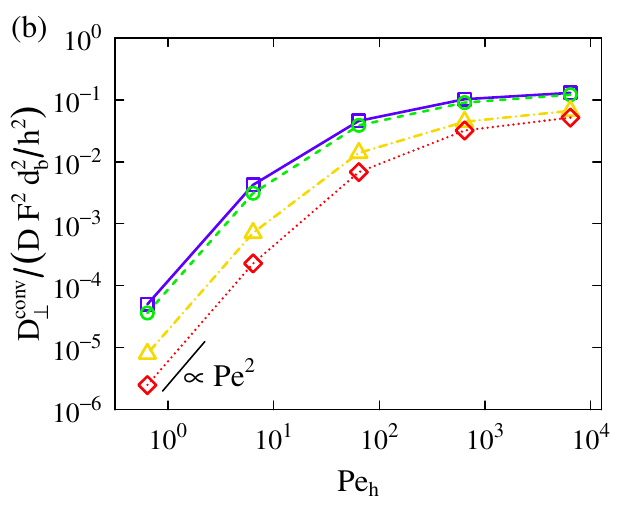}
\caption{
Longitudinal (a) and transverse (b) components of $\m{D}^{\mathsf{conv}}$ as a function of the P\'eclet number \minor{based on the lattice spacing ($\Pe_h = U h / D$)} for ordered arrays in various flow regimes at small volume fraction ($\phi = 0.2$~\%).
\minor{The normalizations of $D^{\mathsf{conv}}_{\parallel,\perp}$ are those suggested by the asymptotic analysis (identical to those used in \cref{fig:analytical_solution_vs_Pe}), and} 
$F$ is given by \cref{eq:def_F_with_U}.
The lines are drawn to guide the eyes.
\minor{Note that in (a), $D^{\mathsf{conv}}_\parallel$ is compensated by $\Pe^2$.}
}
\label{fig:dilute_regimes_Deff_conv_i_vs_Pe}
\end{figure}

We examine the case of suspensions at low (but not vanishing) volume fraction in order to approach the dilute limit assumption, and to focus on the sole effect of inertia.
The longitudinal and transverse components of the convective contribution to the effective diffusivity have been computed \minor{for $h/d_b = 6.4$, which corresponds to a gas volume fraction of} $\phi = 0.2$~\% (the smallest volume fraction accessible with the method and facilities used), for each of the four flow regimes listed in \cref{tab:cases_parameters_sublist}, and 
\minor{$\Pe$} has been varied from $10^{-1}$ to $10^{3}$.
The results are shown in \cref{fig:dilute_regimes_Deff_conv_i_vs_Pe} \minor{as a function of $\Pe_h$, the P\'eclet number based on the lattice spacing, which is the parameter governing the transition between the two asymptotic limits (see \cref{subsec:asymptotic_analysis} and \cref{fig:analytical_solution_vs_Pe}).}
The different colors, symbols and line styles depict the different flow regimes (the lines are drawn to guide the eyes).
Qualitatively, \cref{fig:dilute_regimes_Deff_conv_i_vs_Pe} bears a striking resemblance to \cref{fig:analytical_solution_vs_Pe}, even for case C (skirted bubbles)\minor{: analysis and simulations yield similar dependences of $D^{\mathsf{conv}}_{\parallel,\perp}$ on $\Pe$ and qualitatively comparable effects of increasing $\Re$}. 
\minor{At low P\'eclet number ($\Pe_h \lesssim 10^{1}$), dispersion occurs primarily by molecular diffusion and convective mixing grows quadratically with $\Pe$ in both the longitudinal and the transverse directions ($D^{\mathsf{conv}}_{\parallel,\perp} \propto D \Pe^2$).
At high P\'eclet number ($\Pe_h \gtrsim 10^{3}$), Taylor dispersion is the dominant process, with very efficient mixing in the flow direction ($D^{\mathsf{conv}}_\parallel \propto D \Pe^2$) and negligible mixing in the transverse one ($D^{\mathsf{conv}}_\perp \propto D$).}
\minor{Inertial effects and bubble deformation only affect the proportionality constants, rather weakly for $D^{\mathsf{conv}}_\parallel$ but substantially for $D^{\mathsf{conv}}_\perp$.}

\begin{figure}
\centering
\includegraphics{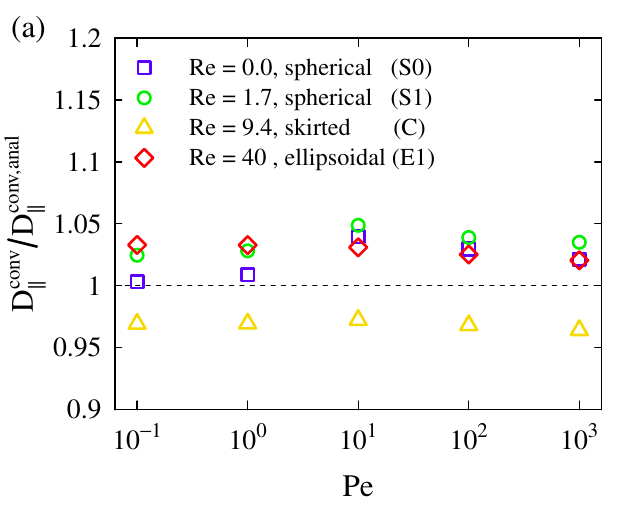}
\includegraphics{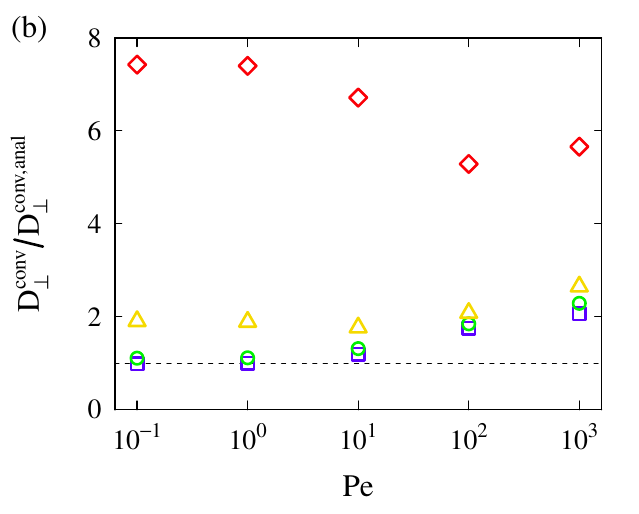}
\caption{
Numerical solution $\m{D}^{\mathsf{conv}}$ divided by the analytical solution $\m{D}^{\mathsf{conv,anal}}$ as a function of the P\'eclet number \minor{based on the bubble diameter ($\Pe = U d_b / D$)} for ordered arrays in various flow regimes at small volume fraction ($\phi = 0.2$~\%): longitudinal (a) and transverse (b) components.
$\m{D}^{\mathsf{conv,anal}}$ is given by \cref{eq:Deff_analytical}.
}
\label{fig:dilute_regimes_Deff_conv_i_num_vs_anal_vs_Pe}
\end{figure}

\begin{figure}
\centering
\includegraphics[width=0.77\linewidth]{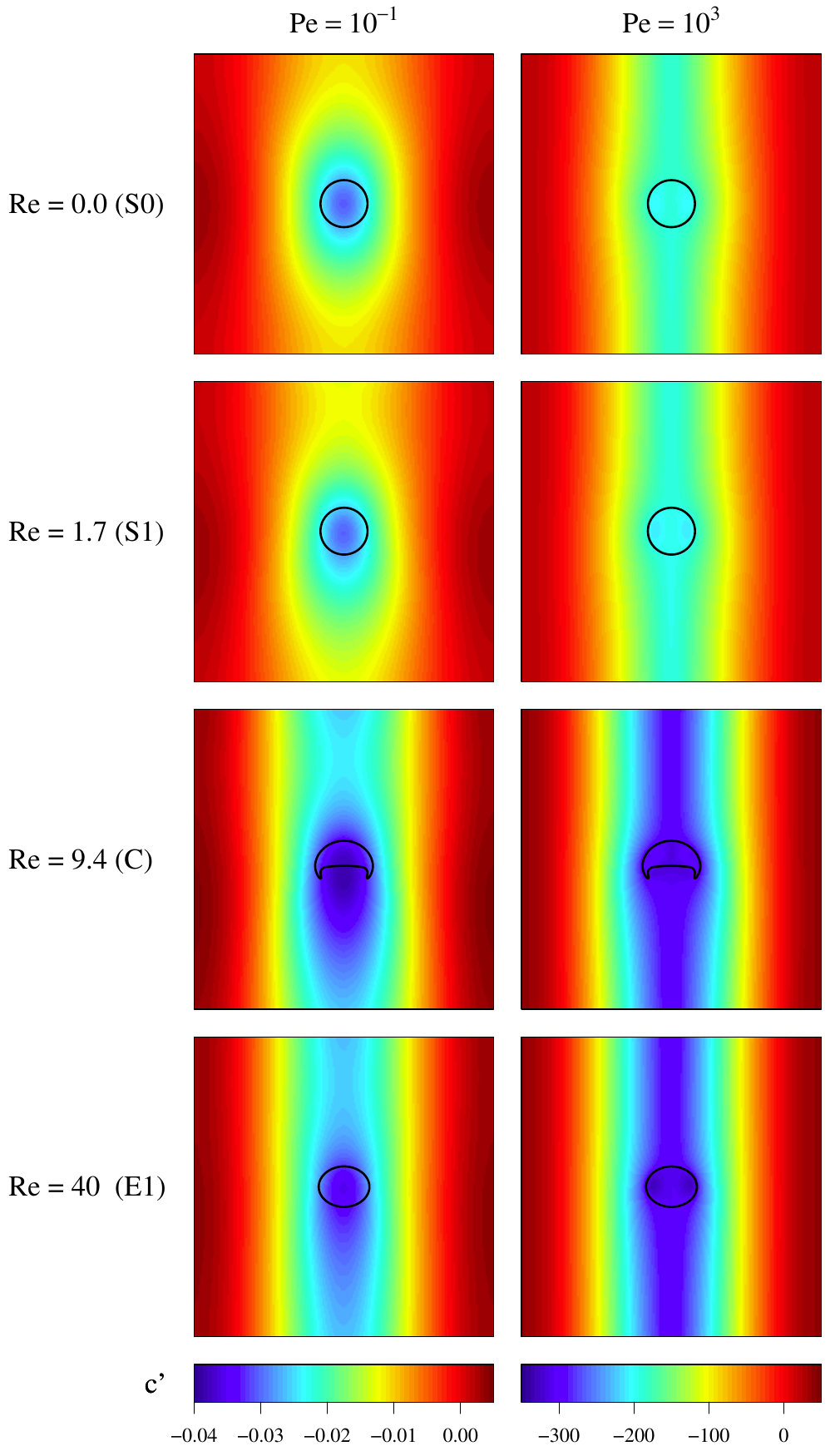}
\caption{
Scalar fluctuation field $c'$ \minor{associated with} $D^{\mathsf{conv}}_\parallel$, shown in a vertical symmetry plane passing through the center of a bubble, for ordered arrays in various flow regimes at $\Pe = 10^{-1}$ (left) and $\Pe = 10^{3}$ (right).
The imposed scalar field $\bar{c}$ increases linearly within the cell from bottom to top ($\phi = 0.2$~\%, the entire cell is shown, and gravity is pointing downward).
}
\label{fig:scalar_fields_2D_plane_by_col_dilute_inertial_effects_z}
\end{figure}

\begin{figure}
\centering
\includegraphics[width=0.77\linewidth]{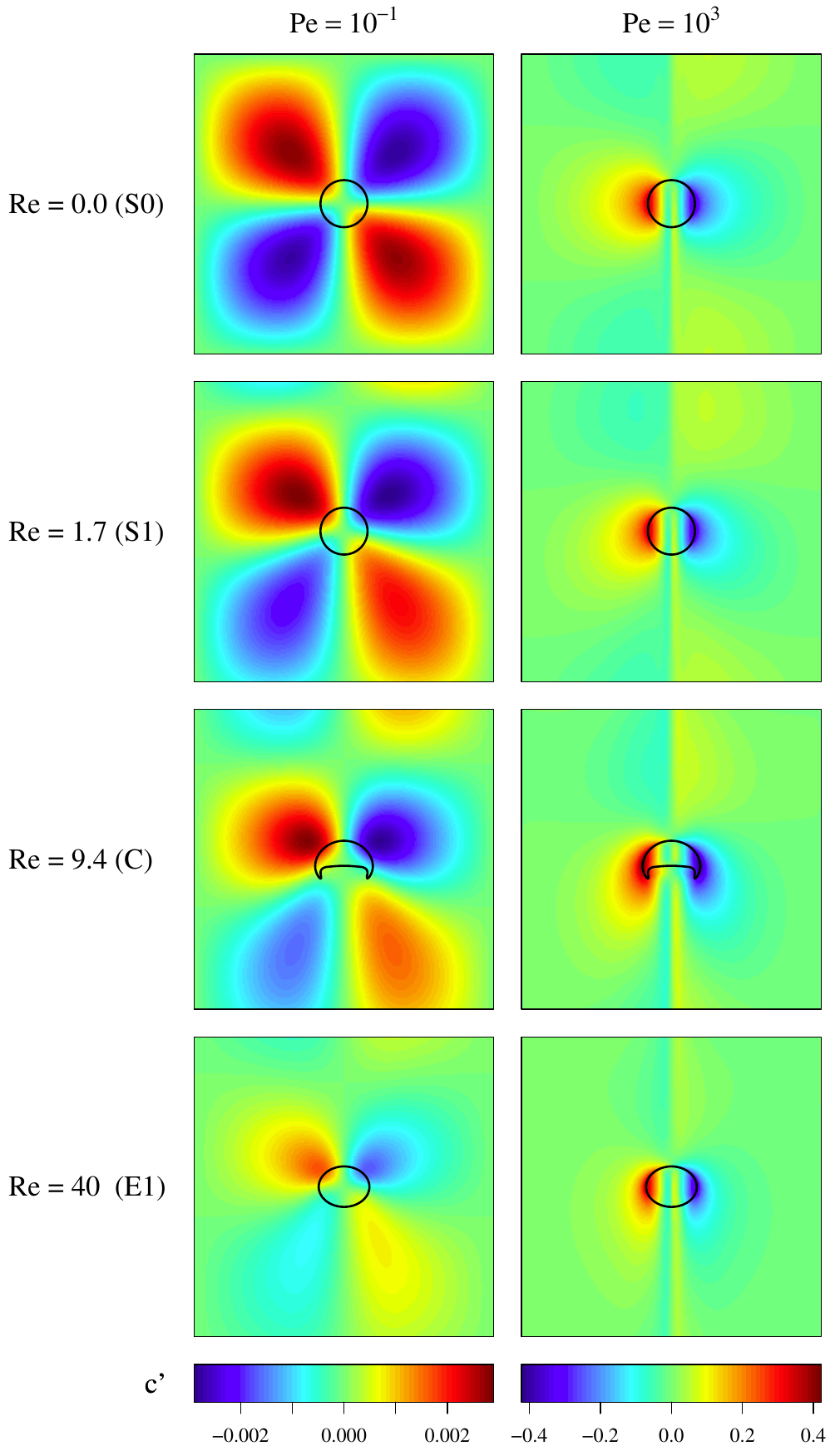}
\caption{
Scalar fluctuation field $c'$ \minor{associated with} $D^{\mathsf{conv}}_\perp$, shown in a vertical symmetry plane passing through the center of a bubble, for ordered arrays in various flow regimes at $\Pe = 10^{-1}$ (left) and $\Pe = 10^{3}$ (right).
The imposed scalar field $\bar{c}$ increases linearly within the cell from left to right ($\phi = 0.2$~\%, the entire cell is shown, and gravity is pointing downward).
}
\label{fig:scalar_fields_2D_plane_by_col_dilute_inertial_effects_x}
\end{figure}

To allow a quantitative comparison between the DNS and the analysis, we present in \cref{fig:dilute_regimes_Deff_conv_i_num_vs_anal_vs_Pe} the ratio of $D^{\mathsf{conv}}_{\parallel,\perp}$ to $D^{\mathsf{conv,anal}}_{\parallel,\perp}$ where $D^{\mathsf{conv,anal}}_{\parallel,\perp}$ is given by \cref{eq:Deff_analytical} with $F$ computed directly from its definition \cref{eq:def_F_with_U}.
\minor{As the range of validity of the analysis is defined in terms of $\Pe$ ($\Pe \ll 1$, with $\Pe$ the P\'eclet number based on the bubble diameter), the data are presented here as a function of $\Pe$ rather than $\Pe_h$.}
For the longitudinal component, the numerical solution does not deviate by more than 5~\% from the theoretical prediction, as can be seen from \cref{fig:dilute_regimes_Deff_conv_i_num_vs_anal_vs_Pe}(a).
\minor{The fact that the low-$\Pe$, Oseen-flow analysis yields accurate predictions for $D^{\mathsf{conv}}_\parallel$ at $\Pe=10^3$ and $\Re=\O(10)$ is not surprising, as the behavior of $D^{\mathsf{conv}}_\parallel / (D F^2 \Pe^2)$ is rather insensitive to both the flow regime and the P\'eclet number (as shown in Fig. \ref{fig:analytical_solution_vs_Pe}(a) and \ref{fig:dilute_regimes_Deff_conv_i_vs_Pe}(a), this quantity does not vary more than 15\% for the cases studied).}
We conclude that, at small volume fraction, \minor{$D^{\mathsf{conv}}_\parallel$ can be predicted within $\pm 5$~\% from \cref{eq:Deff_analytical}} at any P\'eclet number up to $10^3$ and any Reynolds number up to 40, even when the bubbles are strongly deformed.
\minor{For the transverse component, the asymptotic analysis underpredicts} the value of $D^{\mathsf{conv}}_\perp$ at high P\'eclet number, even for $\Re \lesssim 1$.
As a consequence, $D^{\mathsf{conv}}_\perp$ cannot be \minor{accurately} estimated from our analytical solution when the assumptions underlying its derivation are not satisfied.
\minor{It must be kept in mind though that} this component varies much more than the longitudinal one between the regimes of small and large P\'eclet numbers, and is much more sensitive to the flow regime ($\Re$, shape), which means that its value is more difficult to predict. 
\minor{In all, it is worth stressing that the asymptotic analysis yields the correct qualitative behavior and order of magnitude for $D^{\mathsf{conv}}_\perp$ at least up to $\Pe = 10^3$ and $\Re \approx 10$, even for strongly deformed bubbles.}
\minor{Finally, we emphasize} that we found $D^{\mathsf{conv}}_\parallel/D^{\mathsf{conv}}_\perp \gtrsim 10^2$, so the most important component of the effective diffusivity tensor is the longitudinal one, except in situations where there is no longitudinal component of the gradient of the scalar on the macroscale.

To illustrate the dispersion regimes at low and high P\'eclet number, we present in \cref{fig:scalar_fields_2D_plane_by_col_dilute_inertial_effects_z} and \cref{fig:scalar_fields_2D_plane_by_col_dilute_inertial_effects_x} visualizations of the scalar fluctuation field $c'$ used to compute $D^{\mathsf{conv}}_\parallel$ and $D^{\mathsf{conv}}_\perp$, respectively.
In each of these figures, the field of $c'$ is represented for each flow regime in a vertical symmetry plane passing through the center of a bubble for $\Pe = 10^{-1}$ (left) and $\Pe = 10^{3}$ (right), and the Reynolds number increases from top to bottom.
The field of $c'$ associated \minor{with} $D^{\mathsf{conv}}_\parallel$, shown in \cref{fig:scalar_fields_2D_plane_by_col_dilute_inertial_effects_z}, exhibits similar features at low and high $\Pe$.
In contrast, the field of $c'$ associated \minor{with} $D^{\mathsf{conv}}_\perp$, represented in \cref{fig:scalar_fields_2D_plane_by_col_dilute_inertial_effects_x}, is qualitatively different in these two limits.
This illustrates qualitatively why the regimes at low and high $\Pe$ are similar for $D^{\mathsf{conv}}_\parallel$ ($D^{\mathsf{conv}}_\parallel \propto \Pe^2$), whereas the scaling laws identified for $D^{\mathsf{conv}}_\perp$ are different in both limits (see \cref{fig:dilute_regimes_Deff_conv_i_vs_Pe}).
In addition, the Reynolds number and the bubble shape affect the fore-and-aft symmetry and the details of $c'$, but not its essential features, which results in quantitative but not qualitative effects on $D^{\mathsf{conv}}_{\parallel}$ and $D^{\mathsf{conv}}_{\perp}$.

\section{Freely evolving suspensions}
\label{sec:free_arrays}

\begin{figure}
\centering
\includegraphics{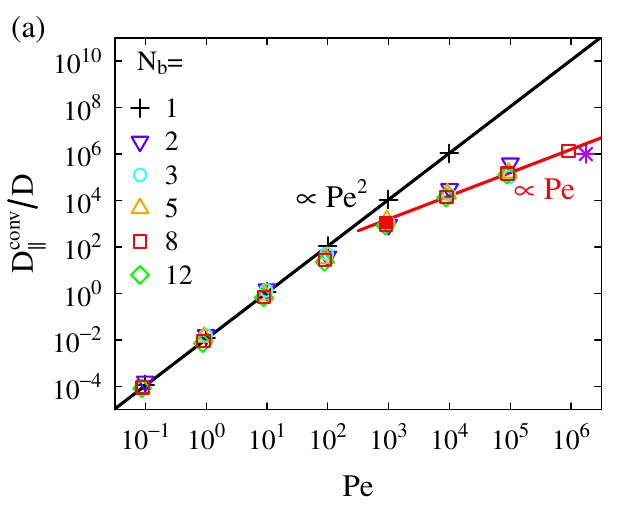}
\includegraphics{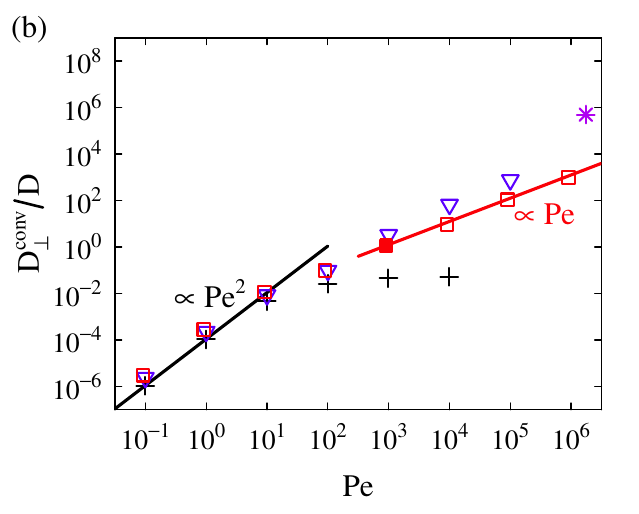}
\caption{
Longitudinal (a) and transverse (b) components of $\m{D}^{\mathsf{conv}}$ as a function of the P\'eclet number for various numbers of free bubbles $N_b$ in the unit cell ($N_b = 1$ corresponds to an ordered array).
Symbols other than purple stars: DNS ($\Re \approx 30$, $\phi = 2.4$~\%); purple stars: experimental data of \textcite{Almeras2015} ($\Re \approx 700$, $\phi \approx 2.4$~\%). A spatial resolution of $d_b/\Delta x = 20$ was used for $N_b > 1$, the effect of increasing resolution to $d_b/\Delta x = 30$ is illustrated by the filled red squares for $N_b = 8$ and $\Pe \approx 10^3$ ($d_b$ is the bubble volume-equivalent diameter and $\Delta x$ is the grid spacing).
}
\label{fig:Dii_vs_Pe_for_various_Nb_heq28}
\end{figure}

We examine in this section scalar mixing in freely evolving suspensions as represented by the periodic repetition of a unit cell containing several independent bubbles (``free arrays'').
Our objective here is threefold: (i) to investigate the effective diffusivity of freely evolving suspensions at small and high P\'eclet numbers, (ii) to compare and contrast these results with those obtained in ordered systems, and (iii) to evaluate the effect of the system size (number of bubbles in a unit cell, $N_b$). 

For that purpose, we considered a single flow regime (ellipsoidal bubbles at $\Re = \O(10)$, corresponding to case E1 in \cref{tab:cases_parameters_sublist}) at intermediate volume fraction ($\phi = 2.4$~\%) and explored the effect of varying the number of free bubbles $N_b$ on the dependence of $\m{D}^{\mathsf{conv}}$ on the P\'eclet number. 
Due to the multiplicity of simulations involved and to their duration (typically several months on 64 cores), only a few different values of $N_b$ belonging to a rather limited range have been considered (namely $N_b = \{2,3,5,8,12\}$ in the simulations for the determination of $D^{\mathsf{conv}}_\parallel$, and $N_b=\{2,8\}$ in those for $D^{\mathsf{conv}}_\perp$). 
For the same reason, investigations of the effects of volume fraction and flow regime could not be undertaken.

The longitudinal and transverse components of $\m{D}^{\mathsf{conv}}$ are plotted in \cref{fig:Dii_vs_Pe_for_various_Nb_heq28} as a function of the P\'eclet number for various values of $N_b$. Note that a very wide range of P\'eclet numbers is considered.
Convergence of $D^{\mathsf{conv}}_{\parallel}$ with the system size is very fast: the values of $D^{\mathsf{conv}}_{\parallel}$ are essentially independent of the number of free bubbles for $2 \leqslant N_b \leqslant 12$ at all P\'eclet numbers.
This suggests that $D^{\mathsf{conv}}_{\parallel}$ is independent of the system size $N_b$, although this would need to be confirmed by considering larger values of this parameter.
Our data for $D^{\mathsf{conv}}_{\perp}$ suggest that convergence with $N_b$ is slower for this quantity, especially at high P\'eclet number, although conclusions can hardly be drawn on this point due to the few values of $N_b$ considered.

\begin{figure}
\centering
\includegraphics[width=0.4\linewidth]{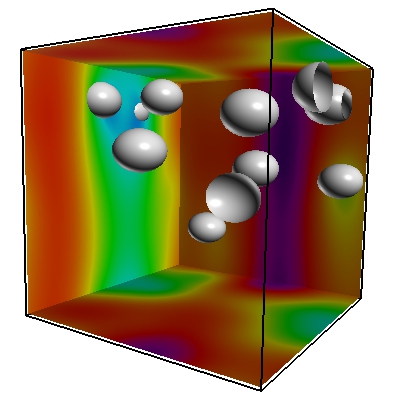}
\includegraphics[width=0.4\linewidth]{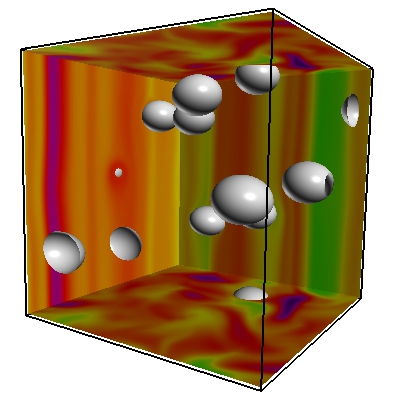}
\caption{Instantaneous scalar fluctuation field $c'$ associated \minor{with} $D^{\mathsf{conv}}_\parallel$ for a free array of 8 bubbles, at $\Pe = 10^{-1}$ (left) and $\Pe = 10^{6}$ (right).
The gradient of $\bar{c}$ is vertical (the entire cell is shown, and gravity is pointing downward).
}
\label{fig:scalar_field_Dzz}
\end{figure}

\begin{figure}
\centering
\includegraphics[width=0.4\linewidth]{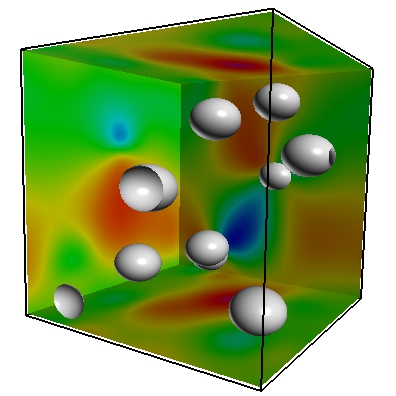}
\includegraphics[width=0.4\linewidth]{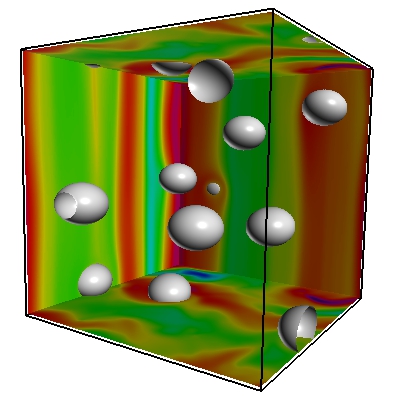}
\caption{Instantaneous scalar fluctuation field $c'$ associated \minor{with} $D^{\mathsf{conv}}_\perp$ for a free array of 8 bubbles, at $\Pe = 10^{-1}$ (left) and $\Pe = 10^{6}$ (right).
The gradient of $\bar{c}$ is horizontal (the entire cell is shown, and gravity is pointing downward).
}
\label{fig:scalar_field_Dxx}
\end{figure}

We first examine the dependence of $\m{D}^{\mathsf{conv}}$ on the P\'eclet number in free arrays of bubbles ($N_b>1$).
At small $\Pe$, $D^{\mathsf{conv}}_{\parallel,\perp} \propto D \Pe^2$, whereas at high $\Pe$, $D^{\mathsf{conv}}_{\parallel,\perp} \propto D \Pe = U d_b$.
Note that the scaling at high $\Pe$ is expected from a simple dimensional analysis in a convection-dominated regime where diffusion plays no role.
This regime corresponds to the ``mechanical dispersion'' regime in \textcite{Koch1985}.
The different dispersion regimes at low and high $\Pe$ can also be identified from the features of the scalar fluctuation field $c'$.
Instantaneous snapshots of $c'$ associated \minor{with} $D^{\mathsf{conv}}_\parallel$ and $D^{\mathsf{conv}}_\perp$ are shown in \cref{fig:scalar_field_Dzz} and \cref{fig:scalar_field_Dxx}, respectively, for an array of 8 free bubbles at $\Pe = 10^{-1}$ (left) and at $\Pe = 10^6$ (right).
For a given component, the isocontours of $c'$ follow markedly different patterns at low and high $\Pe$.

We now compare these results with those obtained for ordered arrays (black crosses in \cref{fig:Dii_vs_Pe_for_various_Nb_heq28}) and discuss the effect of the microstructure.
At small $\Pe$, $D^{\mathsf{conv}}_\parallel$ and $D^{\mathsf{conv}}_\perp$ grow quadratically with $\Pe$ in both free and ordered arrays.
This scaling was also obtained by \textcite{Koch1985} for low-$\Pe$ dispersion in porous media with random microstructure (albeit in the Stokes flow limit).
Since in the low-$\Pe$ regime, diffusion by the random motion of molecules is much faster than convection by the flow, the microstructure has only a quantitative incidence on $\m{D}^{\mathsf{conv}}$, and dispersion is qualitatively identical in ordered and freely evolving suspensions.
Note that similar features in the spatial distribution of $c'$ can be identified in ordered and free arrays at low $\Pe$ (see tubular structures in the left side of \cref{fig:scalar_fields_2D_plane_by_col_dilute_inertial_effects_z,fig:scalar_field_Dzz} for $D^{\mathsf{conv}}_\parallel$, and
\minor{quadrupolar} ones in the left side of \cref{fig:scalar_fields_2D_plane_by_col_dilute_inertial_effects_x,fig:scalar_field_Dxx} for $D^{\mathsf{conv}}_\perp$).
\minor{We however emphasize that precise \emph{quantitative} agreement between the results for one and for many bubbles at low P\'eclet number in \cref{fig:Dii_vs_Pe_for_various_Nb_heq28} is not expected, as the flows and the microstructures in the two systems are different \parencite{Bunner2002a,Loisy2017b}.}

\minor{At high $\Pe$, the Taylor dispersion scaling obtained in ordered arrays is replaced, in both directions, by a scaling similar to the one characterizing mechanical dispersion, as soon as the relative motion between bubbles is allowed. In this regime, the transverse dispersion is indeed governed by mechanical dispersion. Irrespective of the value of $\Pe$, any Taylor dispersion in the vertical direction is limited by transverse \minor{diffusion} or dispersion, the latter becoming more \minor{significant} at large $\Pe$. This results in a scaling similar to that of mechanical dispersion in the longitudinal direction as well, such that a distinction between these two mechanisms (pure mechanical dispersion, or Taylor dispersion limited by transverse mechanical one) cannot be made.}
Incidentally, mechanical dispersion is also obtained at high $\Pe$ in random media in Stokes flow conditions \parencite{Koch1985}.
Although the microstructure of the present bubbly suspensions has not been evaluated quantitatively, visual inspection and prior results on their dynamics \parencite{Loisy2017b} showed that it is not random, but rather characterized by a certain ``organization''.
Despite the fact that freely evolving suspensions resemble ordered ones with respect to their dynamics, scalar dispersion is extremely sensitive to the \emph{presence} of disorder, and is fundamentally different in perfectly ordered and weakly disordered suspensions at high P\'eclet number.
It does not, however, seem to be sensitive to the \emph{degree} of disorder, as suggested by the fact that the same scalings with $\Pe$ are obtained for random porous media and weakly disordered suspensions.
We stress that this last statement is purely speculative, and would require a quantitative study of the effect of the microstructure to be confirmed.

We finally attempt a comparison of our results with the experimental data of \textcite{Almeras2015}, who measured the effective diffusivity of a homogeneous swarm of high-Reynolds-number rising bubbles at $\Pe \approx 1.75 \times 10^6$ for gas volume fractions ranging from 1~\% to 13~\%. 
It is important to stress that in these experiments, $\Re \approx 700$, whereas in the simulations, $\Re \approx 30$, so the comparison is only indicative.
Interpolation (by eye) of their data at $\phi \approx 2.4$~\% (figure 10 in their paper) yields $D^{\mathsf{eff}}_\parallel/D = 1 \times 10^6$ and $D^{\mathsf{eff}}_\perp/D = 5 \times 10^5$.
These experimental values are represented by purple stars in \cref{fig:Dii_vs_Pe_for_various_Nb_heq28}.
Note that at such high P\'eclet number, the dominant contribution to $\m{D}^{\mathsf{eff}}$ is due to $\m{D}^{\mathsf{conv}}$, so it seems reasonable to assume that these are equivalent.
The order of magnitude of $D^{\mathsf{eff}}_\parallel/D$ is comparable in the experiment and in the simulation, whereas $D^{\mathsf{eff}}_\perp/D$ is much higher in the experiment. 
This difference can be explained from the different properties \major{of the numerical and experimental flows considered: partition coefficient (the dye concentration in the gas is presumably zero in the experiments from \textcite{Almeras2015}),
diffusivity ratio, and} 
bubble-induced liquid agitation in the horizontal direction.
In our simulations of free arrays at moderate $\Re$, the bubbles were \minor{indeed} observed to rise along nearly straight vertical lines, and the anisotropy ratio characterizing the liquid velocity variance, $2 \avg{u'_3 u'_3} / \avg{u'_1 u'_1 + u'_2 u'_2}$, is approximately 8 (for $N_b = 8$), whereas in the experiment at high $\Re$, the bubble motion is fully three-dimensional, and the anisotropy ratio is approximately 2.
Finally, as only one value of the P\'eclet number was considered in the experiments of \textcite{Almeras2015}, no comparison of their data with our results can be offered regarding the dependence of the effective diffusivity on the P\'eclet number.

\section{Conclusions}
\label{sec:conclusion}

In this study we investigated scalar dispersion in homogeneous bubbly suspensions as described by an effective diffusivity tensor.
The longitudinal and transverse components of the convective contribution to the effective diffusivity, denoted $D^{\mathsf{conv}}_{\parallel}$ and $D^{\mathsf{conv}}_{\perp}$, respectively, have been computed for bubbly suspensions in various flow regimes.
This convective contribution is that associated with bubble-induced agitation, and is the dominant contribution to the effective diffusivity in commonly encountered bubbly flows.

The dispersion theory of \textcite{Koch1989} indicates that convective mixing mechanisms in ordered suspensions in Stokes-flow conditions differ at low and high P\'eclet numbers.
According to this theory, when the bulk flow is aligned with a primary axis of a simple cubic lattice of spheres, convectively enhanced dispersion is expected at low P\'eclet number, whereas Taylor dispersion should dominate at high P\'eclet number.
In the present study, we have extended this theory to account for weak inertial effects, and we have shown that these two dispersion regimes are qualitatively unchanged in the presence of (weak) inertia.
This result has been confirmed by direct numerical simulations for values of the Reynolds number ranging from vanishingly small to moderate.
In all investigated cases, $D^{\mathsf{conv}}_{\parallel}$ was found to be significantly larger than $D^{\mathsf{conv}}_{\perp}$, and theoretical predictions have been shown to yield 
\minor{the correct qualitative behaviour and order of magnitude of both $D^{\mathsf{conv}}_{\parallel}$ and $D^{\mathsf{conv}}_{\perp}$} in a variety of flow regimes (spherical to strongly deformed bubbles with Reynolds numbers 
\minor{up to 10)} at small volume fraction.

Direct numerical simulations of scalar transport in freely evolving bubbly suspensions, as represented by free arrays of bubbles, have been carried out for a wide range of P\'eclet numbers, and the effect of introducing additional degrees of freedom in the system has been evaluated.
At low P\'eclet number, dispersion in free arrays is convectively enhanced, as in ordered ones.
\minor{At high P\'eclet number, in freely evolving suspensions wherein at least two bubbles are present in a unit cell, the longitudinal component of the effective diﬀusivity exhibits a scaling that is similar to that characterizing mechanical dispersion. This suggests that the limiting role of molecular diffusion to Taylor dispersion is taken over by mechanical
dispersion, or that mechanical dispersion itself dominates.} 
Besides, the effective diffusivity seems to be weakly sensitive to the number of bubbles present in a unit cell.
This last assertion requires more thorough investigations to be confirmed, but is encouraging regarding the possibility of computing the effective diffusivity of homogeneous bubbly flows from direct numerical simulations of systems of relatively small size.
This would allow in particular a thorough investigation of the roles played by the volume fraction and the flow regime, which could not be undertaken as part of the present study.

The results presented in this paper are restricted to bubbles having the same diffusivity as that of the surrounding liquid, and to scalar fields that are continuous across the interface\minor{, and therefore cannot be straightforwardly compared to those obtained in real bubbly flows}.
A jump in the scalar field, which represents the difference in solubilities given by Henry's law in the context of chemical species transport, as well as a difference in diffusivities, would introduce a diffusive contribution to the effective diffusivity tensor \cref{eq:Deff_expr} in addition to the convective one considered in this study.
The present results show the convective contribution at large P\'{e}clet numbers and modest volume fraction to be substantially larger than the diffusive contribution from nonequal diffusivities or solubilities \citep{Maxwell1873book,Jeffrey1973,Koch1985}.
A difference in diffusivities or solubilities would however also have some indirect effect on the convective contribution, which magnitude should be investigated in the future.

Besides the effective diffusivity, another quantity of practical importance is the rate of interfacial scalar transport in the presence of an average scalar gradient between the disperse phase and the bulk.
Heat and mass exchanges across phase boundaries are traditionally expressed as dimensionless transfer coefficients called the Nusselt and the Sherwood numbers, respectively.
Their functional dependences on suspension properties, in particular the volume fraction, have been the subject of analytical \parencite{Acrivos1980b}, numerical \parencite{Aboulhasanzadeh2014}, and experimental \parencite{Colombet2011,Colombet2015} studies.
Formally, the Nusselt and the Sherwood numbers are closure coefficients for the conditionally averaged scalar transport equation, where the conditional average is defined as an ensemble average over the subset of realizations wherein a particulate is present at a given position.
Less formally, the Nusselt and Sherwood numbers are related to a ``mesoscale'' description of scalar transfer between the two phases, whereas the effective diffusivity is associated with a ``macroscale'' description of scalar transport through a two-phase mixture seen as a continuum.
They correspond to different closure problems, and one cannot be inferred from the other.
\major{Nevertheless, the present work will be primarily important for mass transfer processes in bubbly flows that are liquid-phase controlled. This is because then the mixture concentration distribution is key, whereas if it is gas-phase controlled, the concentration in the liquid will be almost uniform and one is primarily concerned by the circumstances inside each bubble.}

\acknowledgments{This work benefited from the financial support of the French research agency (grant ANR-12-BS09-0011), and was performed using the HPC resources provided by GENCI-CINES and GENCI-IDRIS (grant x20162b6893), PSMN (\'Ecole Normale Sup\'erieure de Lyon), P2CHPD (Universit\'e Claude Bernard Lyon 1) and PMCS2I (\'Ecole Centrale de Lyon).
}

\section*{Appendix: Spatial convergence tests}
\label{sec:appendix}

\begin{figure}
\centering
\includegraphics{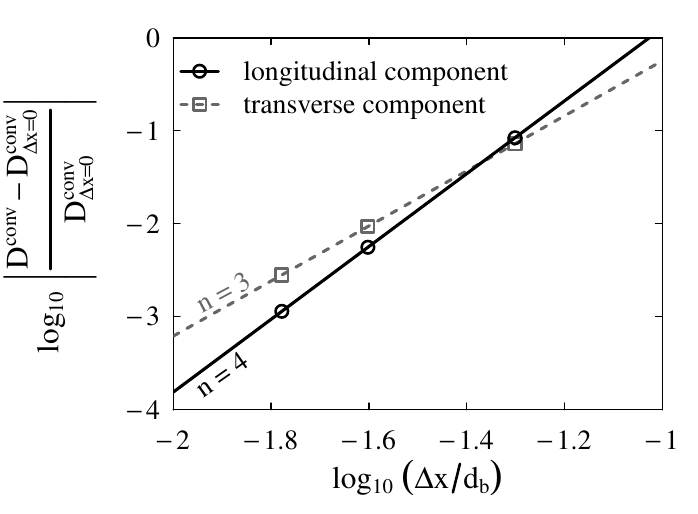}
\caption{
Spatial convergence for an ordered array of bubbles in case E1 at $\Pe = 10^3$: relative error in $D^{\mathsf{conv}}_{\parallel}$ and $D^{\mathsf{conv}}_{\perp}$ as a function of the grid spacing $\Delta x$ ($d_b$ is the bubble volume-equivalent diameter; $D^{\mathsf{conv}}_{\Delta x = 0}$ is extrapolated assuming $D^{\mathsf{conv}} = D^{\mathsf{conv}}_{\Delta x = 0} - k \Delta x^n$, where the values of the three parameters $D^{\mathsf{conv}}_{\Delta x = 0}$, $k$ and $n$ are fitted from numerical data).
}
\label{fig:validation_Deff_spatial_convergence_ET2}
\end{figure}

We present the results of some spatial convergence tests of the algorithm solving the scalar transport equation. The results of similar tests for the algorithm solving the flow are shown in \cite{Loisy2017b,Loisy_thesis}.

The effect of the grid spacing on $D^{\mathsf{conv}}_{\parallel}$ and $D^{\mathsf{conv}}_{\perp}$ has been assessed for case E1 at $\Pe = 10^3$ for one value of the volume fraction ($\phi = 2.4$~\%), in both ordered and free configurations.
For ordered arrays, three different resolutions were tested, namely $d_b/\Delta x = \{20,40,60\}$ with $\Delta x$ the grid spacing.
The results are shown in \cref{fig:validation_Deff_spatial_convergence_ET2}.
The error in the values of $D^{\mathsf{conv}}_{\parallel}$ and $D^{\mathsf{conv}}_{\perp}$ arising from spatial discretization is less than 1~\% when a resolution of 40 grid cells per bubble diameter is used.
This resolution is the same as that used for the simulation of the corresponding bubbly flow in \cite{Loisy2017b}.
In practice, we used for each configuration the same resolution as that selected for the simulation of the corresponding ordered bubbly suspensions (see \cite{Loisy2017b}), namely 60 grid cells per diameter for case C and 40 grid cells per diameter for the other cases.

For free arrays, due to the computational cost of the simulations, only two different resolutions were tested, namely 20 and 30 grid cells per bubble diameter, for an array of 8 bubbles.
Simulations at higher resolution were too expensive to be continued over sufficiently long times to allow a quantitative estimate of the uncertainty.
However the values of $D^{\mathsf{conv}}_{\parallel}$ and $D^{\mathsf{conv}}_{\perp}$ obtained with the finer grid, depicted by filled red squares in \cref{fig:Dii_vs_Pe_for_various_Nb_heq28}, are nearly indistinguishable from those obtained with the coarser grid.
A resolution of 20 grid cells per diameter was therefore concluded to be sufficient for free arrays in view of the present purposes.

For a given case, the same resolution was used for all P\'eclet numbers.
Note that when the gas diffusivity differs from that of the liquid (a situation not considered here but frequently encountered in practice), finer resolutions may be required, as thin scalar boundary layers around the bubbles would then need to be resolved.

\bibliographystyle{jfm}
\bibliography{biblio}

\end{document}